\documentclass{elsart}
\usepackage{graphicx}
\usepackage{amsmath}
\usepackage{amsfonts}
\usepackage{amssymb}
\usepackage{ulem}

\begin{document}
\normalem
\begin{frontmatter}
\title{In-Vivo NMR of Hyperpolarized $^3$He in the Human Lung
at Very Low Magnetic Fields}
\author{Christopher P. Bidinosti\thanksref{t1},}
\author{Jamal Choukeife,}
\author{Pierre-Jean Nacher,}
\author{Genevi\`{e}ve Tastevin}
\address{Laboratoire Kastler Brossel, 24 rue Lhomond, F75231 Paris, France\thanksref
{t2}}
\thanks[t1]{Corresponding author. E-mail:
cpbidino@sfu.ca  Present address: Department of Physics, Simon
Fraser University, Burnaby, Canada V5A 1S6 }
\thanks[t2]{The Laboratoire Kastler Brossel,
member of the D\'{e}partement de Physique de
l'Ecole Normale Sup\'{e}rieure, is UMR 8552 of the CNRS and of Universit\'{e}
Pierre et Marie Curie}
\date{\today}
\begin{abstract}
We present NMR measurements of the diffusion of hyperpolarized
$^3$He in the human lung performed at fields much lower than those
of conventional MRI scanners. The measurements were made on
standing subjects using homebuilt apparatus operating at 3~mT.
O$_2$-limited transverse relaxation ($T_2$ up to 15-35~s) could be
measured in-vivo. Accurate global diffusion measurements have been
performed in-vivo and in a plastic bag; the average apparent
diffusion coefficient (ADC) in-vivo was
$14.2\pm0.6~\mathrm{mm^{2}/s}$, whereas the diffusion coefficient
in the bag ($^3$He diluted in N$_2$) was
$79.5\pm1~\mathrm{mm^{2}/s}$.
  1D ADC mapping with high SNR
 ($\sim$~200~- 300)
 demonstrates the real possibility
of performing quality lung imaging at extremely low fields.
\end{abstract}
\begin{keyword}
Hyperpolarised helium; Lung imaging; Diffusion; ADC.
\end{keyword}
\end{frontmatter}

\section{Introduction}

The first demonstrations of very low field MRI using a hyperpolarized noble
gas were done about four years ago, first in glass and plastic
cells~\cite{tseng}, and subsequently in excised rat lungs~\cite{wong}. These
measurements were performed at $\sim$~2~mT with a homebuilt instrument, and
showed, in convincing fashion, the feasibility of generating MR images with
hyperpolarized gas (HG) in magnetic fields that are otherwise far too weak to
have produced a useful thermal Boltzmann polarization. There are many possible
uses for this technique including medical imaging. In this paper we present a
realization of this idea through the first measurements of the apparent
diffusion coefficient (ADC) of hyperpolarized \thinspace$\mathrm{{^{3}He}}$ in
the human lung at very low field (3~mT).

The very first use of HG to image a biological system was reported in 1994;
here, the excised heart and lungs of a mouse where imaged using hyperpolarized
\thinspace$\mathrm{{^{129}Xe}}$ in a commercial NMR unit operating at 9.4~T
field~\cite{albert}. In-vivo studies on the human lung began in 1996 using
conventional clinical systems operating at 0.8~T~\cite{ebert,bachert} and
1.5~T~\cite{macfall}. However, unlike conventional MRI, the strong field is
not needed here to polarize the gas. Futhermore, if experimental noise is
dominated by electrical losses in the body it scales linearly with
frequency~\cite{hoult}, and because the inductive NMR signal does so as well,
the signal to noise ratio (SNR) is independent of field and there exists no
benefit to performing HG imaging at high fields in that respect. This was
confirmed at 0.1~T by the research group at Orsay using a
Sopha-imaging-Magnetech MRI unit~\cite{darrasse,durand}.

Still, all in-vivo HG studies reported to date have employed
\emph{high field} magnets designed for proton imaging, a fact most
likely due to the reason that researchers have found it convenient
for the moment to simply adopt existing technology. The question
remains, however, at which field strength could HG imaging prove
itself most useful for medical diagnostics and will this warrant
the development of specialized HG scanners? From this point of
view, there are several compelling arguments to investigate very
low field ($<$~0.1~T) HG imaging. We list them here as the
underlying motivation for the work presented in this paper.

a) \emph{Field Independent SNR} - As mentioned, there is no
reduction in SNR at lower fields if RF losses in the body dominate
the noise. However, below some frequency the body becomes
transparent to the RF and the Johnson noise of the detection coils
will subsequently dominate. Coil-dominated noise depends on the
field as $B\mathrm{^{1/4}}$~\cite{hoult}, so one expects the SNR
to vary as $B\mathrm{^{3/4}}$ at very low fields. The frequency at
which this cross-over occurs has yet to be established
experimentally, but it is expected to be larger than the 102~kHz
used in our study here. From this point of view, a larger field
strength may be preferable, however one should note
 that a lower than optimal SNR
will not necessarily preclude operation in the very low field
regime, as there may exist a desirable trade-off with the other
possible benefits.

b) \emph{Reduced Susceptibility Gradients} - Spatial variations in tissue
susceptibility produce local field gradients whose magnitude is proportional
to the applied field. This can be problematic for lung imaging at high fields,
where differences of magnetic susceptibility between blood vessels, lung
tissue and airspaces produce very strong gradients.

c) \emph{Reduced Power Absorption Rate} - RF power absorption
rates decrease with frequency. Therefore, operation at lower
fields will allow the use of new, rapid pulse sequences without
exceeding the safety limit of 4~W/kg of absorbed power. This has
been demonstrated at 0.1~T~\cite{durand} with $\pi$-pulse
repetition times as short as 10~ms.

d) \emph{Novel Detection Methods} - It may be possible to develop innovative
signal detectors that offer improved performance at low field, but would
otherwise be impractical for use at high field. Possible examples include
networks of local detectors, superconducting coils, and SQUID detectors and amplifiers.

e) \emph{Novel Scanner Geometries} - Very low field scanners will not require
cryogenics (to cool superconducting magnets) nor be encumbered by massive
resistive or permanent magnets. This will permit unparalleled flexibility in
scanner design. For example, a less bulky system may allow for measurements to
be made with the subject either standing or supine, which might be of interest
in pulmonary research.

f) \emph{Low Cost, Dedicated Scanners for Pulmonary Research} -
There is strong evidence that the occurrence of certain lung
pathologies, such as asthma and chronic obstructive pulmonary
disease (COPD), is increasing, and that HG imaging could provide
improved diagnosis and assessment of these prevalent
afflictions~\cite{mayo}. Should the usefulness of this technique
become convincingly established, demands for time on existing
conventional scanners could
 increase dramatically. Very low field, HG scanners
should be relatively inexpensive to purchase and operate, and
 as a result, they might one
day find a use in this research area.

Having put forth these proposed benefits,
 we also point out certain challenges that may exist in performing
 very low field HG imaging.  First, as stated above, SNR will
 eventually degrade with reduced field.  As a result,
  operating at too low a field would most likely force one
  to reduce Johnson noise by means of a cryogenic system to cool
   conventional or superconducting
pick-up coils. Another challenge may also arise if one attempts to
make ultrafast diffusion measurements or imaging. In this case,
the strong applied gradients may have longitudinal and transverse
components that approach, or even exceed, the amplitude of the
uniform applied magnetic field, thereby precluding straightforward
operation~\cite{mohoric,norris,durandthesis}.  For example, an
applied gradient of 15~mT/m would result in additional field
components up to $\sim \pm 2$~mT over the region of the lung; to
determine ADC values using such a gradient at very low $B_{0}$
field (such as 3~mT), one would have to abandon the usual high
field approximation and consider a more exact calculation
involving diffusion along the line gradient of the total magnetic
field~\cite{mohoric}.

To move well beyond the point of conjecture and to truly examine
the potential of very low field lung imaging with hyperpolarized
gas, a systematic study of the ideas laid out above must commence.
(Additional demonstrations of HG imaging have recently been
reported at 15~mT~\cite{albertRAT}). The goal of our research here
was to begin this exploration, first, by developing appropriate
apparatus, and second, by performing in-vitro and in-vivo CPMG
measurements~\cite{meiboom} using hyperpolarized
\thinspace$\mathrm{{^{3}He}}$. The results are compared most
extensively with those made at 0.1~T~\cite{darrasse,durand}, as
this is the lowest field at which the experimental parameters
associated with HG lung imaging have previously been tested.

\section{Apparatus}

\subsection{$\mathrm{B_{0}}$ and $\mathrm{B_{1}}$ Coils}

\label{sec:B0B1}
\begin{figure}[tbh]
\centering
\includegraphics[keepaspectratio,width=3.4in,clip= ]{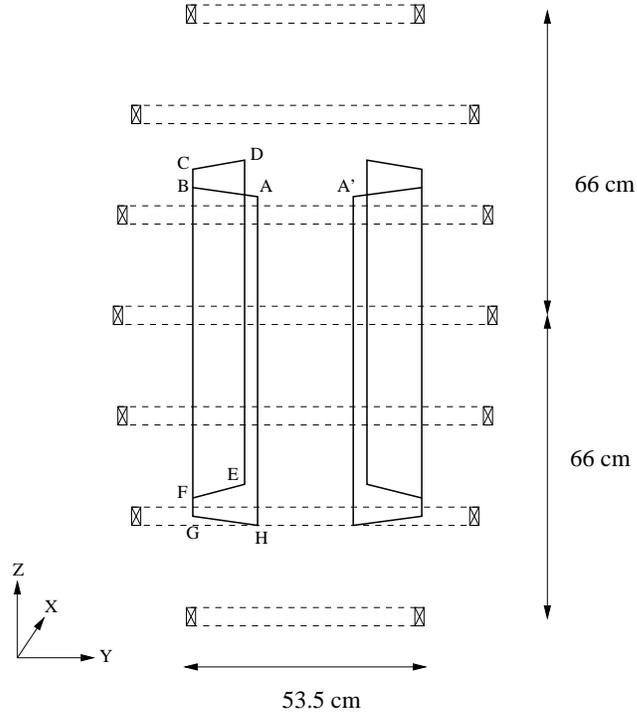}
\caption{Diagram of the $B_{0}$ and $B_{1}$ coils. The coils of
the $B_{0}$ magnet are shown in cross-section: six coils (two each
of mean diameter 78.0, 71.5, and 53.5~cm) are placed symmetrically
(at 22.5, 44.5, and 66.0~cm respectively) about the central coil,
which has a mean diameter of 79.5~cm. For clarity, the $B_{1}$
coils are rendered in 3-D; the corners are labeled alphabetically.
All pertinent dimensions are given in the text.}
\label{fig:mainmag}
\end{figure}

For this first set of measurements, we have employed a
previously made $B_{0}$ magnet. The magnet consists of seven co-axial coils
each of 144~turns of 2~mm diameter copper wire wound on an annular aluminum
frame. The relative orientation of the coils is shown in cross-section in
Figure~\ref{fig:mainmag}. The relative homogeneity of the $B_{0}$ field is only
$\sim 2000$~ppm over a typical lung volume ($\Delta x\Delta
y\Delta z=\mathrm{15\times15\times30~cm^{3}}$). The magnet was
driven by a standard laboratory power supply and could maintain
fields up to 6~mT (for 8.7~A, 940~W) without forced cooling. The
bore of the magnet is along the vertical (Z-direction), and the
entire construction is supported on an aluminum frame (not shown)
with the bottom coil being 75~cm above the floor. This provides
sufficient room for an average sized person to enter from below
and rest standing during measurements.

The subject stands facing the X-direction, with chest centered
within the two ``C-shaped'' coils (shown in
Fig.~\ref{fig:mainmag}) that generate the RF field, $B_{1}$, along
the Y-direction. These coils were wound on a PMMA (poly(methyl
methacrylate))
  frame with 11
outer turns (wound A through H) and 9 inner turns (wound B,C,F,G).
The outer sections ($\mathrm{\overline{AB}}$~=~15.4~cm) make a
$109^{\circ}$ angle with the inner sections
($\mathrm{\overline{BC}}$~=~32.0~cm); the height of the coils is
72~cm and they are separated from one another by a distance
$\mathrm{\overline{AA^{\prime}}}$~=~15.6~cm. The coil was designed
(using home-written software) as an inductor to produce an RF
field homogeneous to within $5\%$ over a typical lung volume.
(This was confirmed by mapping with a small search coil.) To avoid
spurious current paths, copper wire with a relatively thick
insulating layer (1.4~mm diameter wire covered by a 0.7~mm layer
of plastic) was used to keep adjacent windings well separated,
thereby minimizing internal capacitance and maintaining a
sufficiently high self-resonance of 350~kHz. The coil was tuned to
the Larmor frequency with a series capacitor to obtain a
convenient low impedance ($Z_{1}\sim50$~$\Omega$, $Q_{1}=12$). It
was driven by a homebuilt RF amplifier (with APEX PA46 operational
amplifiers) and could produce a $\pi$-pulse at 100~kHz
($B_{0}\mathrm{\sim3~mT}$) in a duration of
$T_{\mathrm{RF}}=0.8$~ms. Gating and dephasing of the RF pulse
were controlled via homebuilt circuitry. The reference frequency
came from the oscillator output of an EG\&G 7265 lock-in
amplifier, also used for NMR signal detection.

\subsection{Detection Coils}

\begin{figure}[tbh]
\centering
\includegraphics[keepaspectratio,width=2.5in,clip= ]{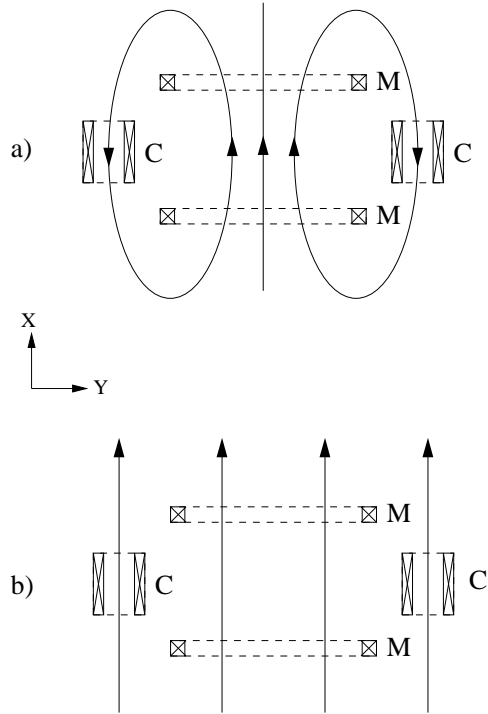}
\caption{Detection coil configuration. The compensation coils (C)
are connected in series opposition with the main coils (M). Field
lines are shown for: a) an internal NMR signal, which is augmented
by the compensation coils, b) an external homogeneous noise
signal, which is nullified by the compensation coils.}
\label{fig:pickup}
\end{figure}

The detection system consists of four rectangular coils: two main coils (each
of ten turns and area $\mathrm{34\times40~cm^{2}}$) at the front and back of
the chest, and two smaller \emph{compensation} coils (each of forty turns but
with $1/4$ the area of the main coils) on either side of the chest. The main
coils are separated by 26~cm, the compensation coils by 50~cm. The coils were
arranged symmetrically about the center of the $B_{1}$ coil, and face the
X-direction perpendicular to the $B_{1}$ field. The compensation coils were
made smaller due to space limitations inside the $B_{0}$ magnet, but have the
same flux (i.e. area $\times$ number of turns) as the main coils. They were
added to the circuit, as sketched in Figure~\ref{fig:pickup}, to increase the
signal to noise ratio: the two sets of coils are connected in series, but with
opposite polarity, which suppresses external noise signals while augmenting
the internal NMR signal. To measure the \thinspace$\mathrm{{^{3}He}}$ NMR signal, the coil
system was tuned globally to the Larmor frequency ($Q=32$), and
the voltage signal was detected using the lock-in amplifier with
differential input.

\begin{figure}[tbh] \centering
\includegraphics[keepaspectratio,width=3.4in,clip= ]{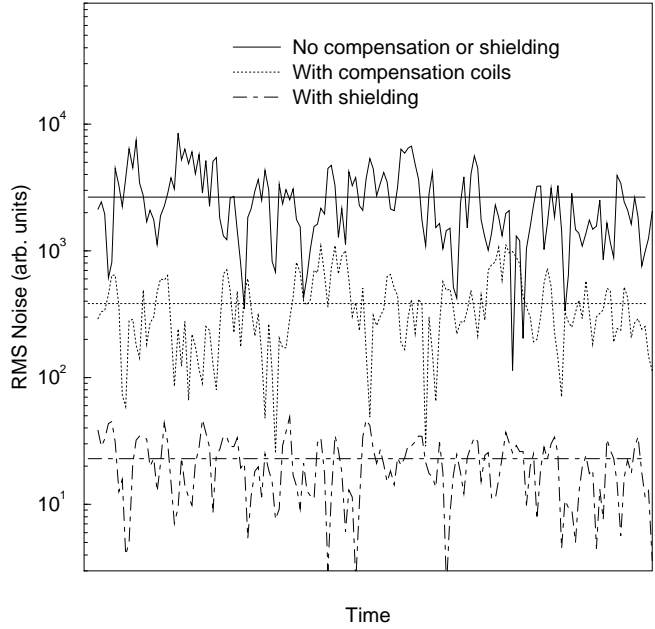}\caption{Time
sample of noise at lock-in output. Horizontal lines give the
average value as calculated over 4 s (for clarity, only 80 ms of
data are shown). With the use of the compensation coils (dotted
line), the background noise (solid line) was reduced by a factor
of 7. A further reduction, by a factor 17, was gained by enclosing
the experiment with copper shielding (dot-dashed line).}
\label{fig:noise}
\end{figure}

Typical noise levels at the lock-in output are shown in
Figure~\ref{fig:noise}.
The use of the compensation coils reduced the background noise by
a factor of 7, which was sufficient to observe the desired NMR signal and make
global measurements of the ADC. However, to obtain the necessary resolution
for regional measurements, the entire experiment eventually had to be enclosed
within a Faraday cage. This brought a further reduction in noise by a factor
of 17 as shown in Figure~\ref{fig:noise}. The shielding (0.5~mm copper sheet)
was effective in reducing external noise sources to levels below that of the
intrinsic noise of the detection system (dominated by the Johnson noise of the
coils). Under these conditions, the SNR was slightly better without the use of
the compensation coils, as the added noise owing to their large impedance was
not offset by their contribution to the signal.

\subsection{Gradient Coil for Diffusion Measurements}

A gradient coil was made from two pairs of anti-Helmholtz coils (two 7-turn
coils separated by 51.6~cm, and two 1-turn coils separated by 20.7~cm) wound
on the $B_{1}$ coil frame. The combination of these coils produces a linear
gradient $dB_{0}/dz=95$\textrm{~}$\mu$T/m per Ampere. Measurements using a
calibrated gradiometer \cite{senaj} showed the gradient field to be
homogeneous to within $5\%$ over the typical lung volume. The gradient coil
was driven by a Kepco 50-8M bipolar power supply with current switching times
$\mathrm{<25~\mu s}$. Eddy-currents generated in the aluminum frame of the
$B_{0}$ magnet have been observed to hamper fast gradient switching by
$\sim15\%,$ and to have a decay time of $\mathrm{800~\mu s}$. Pre-emphasis was
used to compensate for their effect, and the desired gradient is obtained to
$\pm3\%$ of the change after $\mathrm{100~\mu s}$ and to $\pm0.5\%$ of the
change after $\mathrm{300~\mu s.}$

\subsection{Gas Production}

Production of the hyperpolarized \thinspace$\mathrm{{^{3}He}}$ was
done in-house using the technique of metastability exchange
optical pumping~\cite{chupp,tastevin}. The gas was polarized at a
pressure of about 3~mbar, within a homogeneous magnetic field of
0.9~mT, using a 2W ytterbium fibre laser (Keopsys YFL-1083-20).
The pumping was done in a 50~cm long optical cell. The flow of gas
through the cell was controlled via the combination of a flow
regulator and a peristaltic compressor~\cite{compressor}, which is
used to extract the polarized gas from the optical cell and
accumulate it in a storage cell. With flow rates around
2~$\mathrm{cm^{3}}$ of gas at standard pressure per minute, a
polarization of $30-40\%$ could be achieved in the storage cell.
Typically, 40~standard $\mathrm{cm^{3}}$ of helium (1.8~mmol) were
used for each
 measurement.

\section{Method}

When the gas was required for an experiment, it was pumped back via the
peristaltic compressor into a 1~liter plastic bag (a Tedlar gas sampling bag
from Jensen Inert Products) to be carried to the measurement apparatus. High
purity nitrogen was used as a neutral buffer gas to inflate the bag with a
total amount of gas usually of order 0.5 liter. The relaxation time $T_{1}$ of
the gas in the bag is $\sim$~20~minutes, so experiments were designed to
commence soon after this transfer had been made.

The extraction and transport of the gas was done in earth's field
only, the optical pumping field and the NMR field having both been
switched off. Just prior to measurement, and with the bag of gas
inside the apparatus, the $B_{0}$ magnet current was ramped back
up. In-vitro measurements were performed in the bag. In-vivo
measurements were performed, with governmental approval, using
4~healthy adults who were fully informed about the procedures.
 While standing within the detection coils, the
subject exhaled normally, inhaled the gas directly from the bag,
and then further inhaled air to completely fill the lungs. The
subject was then required to hold his breath during the
acquisition time (typically 5~s or less).

All data presented in this paper were taken at 102~kHz
($\sim3$~mT) using CPMG echo sequences~\cite{meiboom}. A
$90_{x}^{\circ}$ RF pulse ($T_{\mathrm{RF}}=0.4$~ms) was followed
by a train of $180_{y}^{\circ}$ RF pulses
($T_{\mathrm{RF}}=0.8$~ms) to refocus the transverse magnetization
at a regular interval $T_{\mathrm{CP}}$. Spatially non-selective
tipping was obtained with a good angle accuracy by switching off
any applied gradient at least $\mathrm{400~\mu s}$ before the
beginning of the RF\ pulses (see App.~\ref{app1}). A PC running
lab-written software was used to manage the RF and gradient pulse
sequencing, as well as to collect data directly from the lock-in
amplifier via an A/D converter.

For a CPMG sequence, the transverse magnetization has a decay time

$T_{2'}$ that can be written as
\begin{equation}
\frac{1}{T_{2'}}=\frac{1}{T_{2}}+\frac{1}{T_{2,\mathrm{diff}}}
\label{eq:t2general}
\end{equation}
The first term on the right hand side is the inherent relaxation
rate of the system, the second term represents additional
relaxation resulting from diffusion in a non-uniform magnetic
field, due to an applied field gradient and/or the magnet
inhomogeneities. For a uniform applied gradient $G$,
$T_{2,\mathrm{diff}}$ can be expressed in the case of free gas
diffusion as
\begin{equation}
T_{2,\mathrm{diff}}=\frac{12}{D(\gamma kGT_{\mathrm{CP}})^{2}}
\label{eq:tdiff}
\end{equation}
where $D$ is the coefficient of diffusion and the factor $k$ corrects for the
time variation of the applied gradient, which is switched off during the RF
pulses. For a rectangular gradient pulse, this correction factor is given as
\begin{equation}
k=\sqrt{\delta^{2}(3T_{\mathrm{CP}}-2\delta)/T_{\mathrm{CP}}^{3}}
\label{eq:Gcorr}
\end{equation}
where $\delta\leq T_{\mathrm{CP}}$ is the duration of the
gradient~\cite{stejskal}.

CPMG measurements were made for different values of
$T_{\mathrm{CP}}$ (3.6 to 90~ms) and applied gradient $G$ (0 to
660~$\mathrm{\mu T/m}$).  For those measurements involving finite
$G$, global or regional values of $D$ were extracted using
 Equation~\ref{eq:tdiff}.  This formula remains valid here, since
 over the length
 $\Delta z \sim \pm 15$~cm of the lung these relatively small applied gradients
  result in additional longitudinal (transverse) field components
 that are at least a factor 30 (60) smaller than the uniform
 3~mT field~\cite{mohoric}.

For in-vitro measurements in the bag, this should almost exactly
correspond to $D_{\mathrm{HeN2}}$, the coefficient for free
diffusion of $\mathrm{{^{3}He}}$ in $\mathrm{{N_{2}}}$ (see
App.~\ref{app2}). For in-vivo measurements, the extracted value of
$D$ was considered an apparent diffusion coefficient (ADC) for the
restricted diffusion of $\mathrm{{^{3}He}}$ in the gas mixture
inside the lung.

\section{Results}

Preliminary measurements of the $B_{0}$ homogeneity were made via
NMR on a sealed cell ($\mathrm{\sim75~cm^{3}}$) of low pressure
\thinspace $\mathrm{{^{3}He}}$ gas. The use of a small set of
receive coils gave excellent SNR ($\sim$~100) despite the low
pressure of the gas (1 to 6~mbar). The gas in the cell was
optically polarized in-situ, which allowed the $B_{0}$ field to be
mapped with relative ease. At each position an FID was recorded
following a $90^{\circ}$ RF pulse; the frequency of the signal
gave the local field strength, whereas $T_{2}^{\ast}$, the decay
time of the FID at $1/e$ of its initial value, gave a measure of
local gradients. Measurements were made at fields of 6~mT and
3~mT. In both cases, frequency measurements conformed to the
theoretically
 calculated field map of the
magnet. Furthermore, values of $T_{2}^{\ast}$ were found to be
twice as long at the lower applied field, which confirmed that
local fields and their gradients were still dominated by the
magnet at this level. To minimize residual gradients, we chose to
operate at 3~mT or less.

\begin{figure}[tbh]
\centering
\includegraphics[keepaspectratio,width=3.4in,clip=]{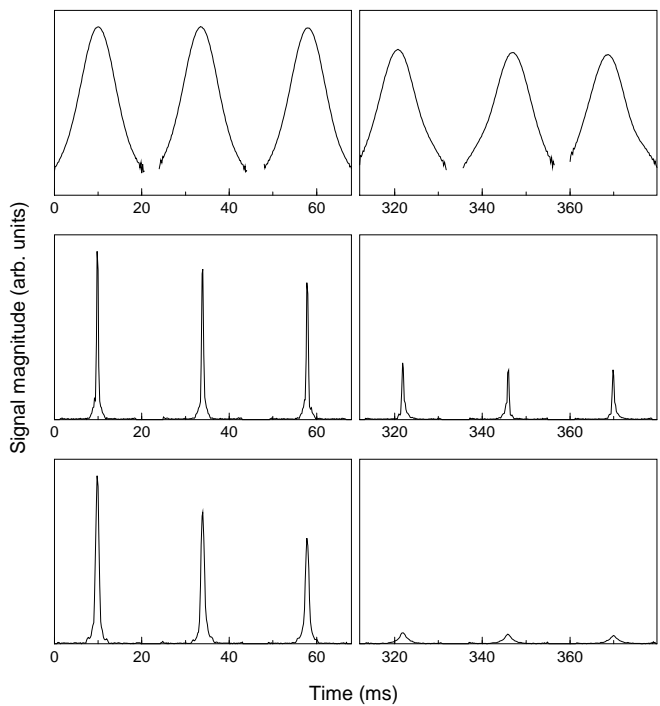}\caption{CPMG
spin echo data with $T_{\mathrm{CP}}=24.8$~ms. Saturation of the
detection system occurs between echos due to the strong RF pulse,
and the corresponding data are not displayed. Top panels: in-vivo,
zero applied gradient. Echo amplitude shows little decay after
300~ms; residual magnet gradients give a
$T_{2}^{\ast}\simeq10$~ms. Middle panels: in-vivo,
$G=380~\mathrm{\mu T/m}$. The strong applied gradient shortens
 $T_{2'}$ and $T_{2}^{\ast}$. Bottom panels:
in-vitro, $G=240~\mathrm{\mu T/m}$. With free diffusion, a shorter
 $T_{2'}$ is obtained despite the lower applied
gradient.} \label{fig:echo}
\end{figure}

To display the operation of the apparatus as well as some of the
characteristics of $\mathrm{{^{3}He}}$ gas NMR, a collection of CPMG spin echo
data is shown in Figure~\ref{fig:echo}. 
These measurements were made with the Faraday cage and exhibit
excellent SNR's of 200 to 300. The repetition rate was
$T_{\mathrm{CP}}=24.8$~ms. The time axes of the graphs are broken
to accentuate the evolution of the echoes. Data shown in the top
panel were taken in-vivo with zero applied gradient, and exhibit
an echo amplitude that decays little over the timescale shown
($T_{2'}\sim 3.3$~s).
 The decoherence time $T_{2}^{\ast}\sim10$~ms (Full Width Half
Maximum), and the corresponding spread in the frequency domain,
 90~Hz FWHM, are
 consistent with the inhomogeneity
of the $B_{0}$ field over this volume. This suggests that tissue
susceptibility gradients contribute little to the relaxation
observed here. Data shown in the middle panel were taken in-vivo
with $G=380~\mathrm{\mu T/m}$. As a result of the strong applied
gradient, the echoes are much more narrow (shorter $T_{2}^{\ast}$)
and show a marked decay in amplitude (shorter $T_{2'}$, now
0.34~s). Data shown in the bottom panel were taken in-vitro with
$G=240~\mathrm{\mu T/m}$; to obtain a decay rate similar to that
which was observed in-vivo, the value of the applied gradient has
to be lowered to compensate for a faster rate of diffusion (see
Eq.~\ref{eq:tdiff}). This result highlights the difference between
the restricted diffusion of $\mathrm{{^{3}He}}$ inside the lung
and the free diffusion it experiences in the bag.

Apart from the slowest decays recorded in-vivo (see Sec 4.1), no
significant deviation from monoexponential behavior was observed
in the diffusion weighted echo decays . To extract values of
$T_{2'}$ from our data, monoexponential fits were made to the
square of the echo amplitudes, or the square of their Fourier
components. The use of power data, rather than the magnitudes,
yields a more reliable result as the latter contains a non-zero
baseline that leads to
 an overestimation (underestimation) of decay times if it is left in
 (subtracted out)~\cite{weerd}.
For extremely long decay times, such as those shown in
Figure~\ref{fig:pipulse}, we found that the results from various
fitting procedures did not vary significantly (this is further
discussed in Sec.~\ref{sec:regional}).

\subsection{Fast Repetition CPMG with no applied gradient}

\begin{figure}[tbh] \centering
\includegraphics[keepaspectratio,width=3.4in,clip= ]{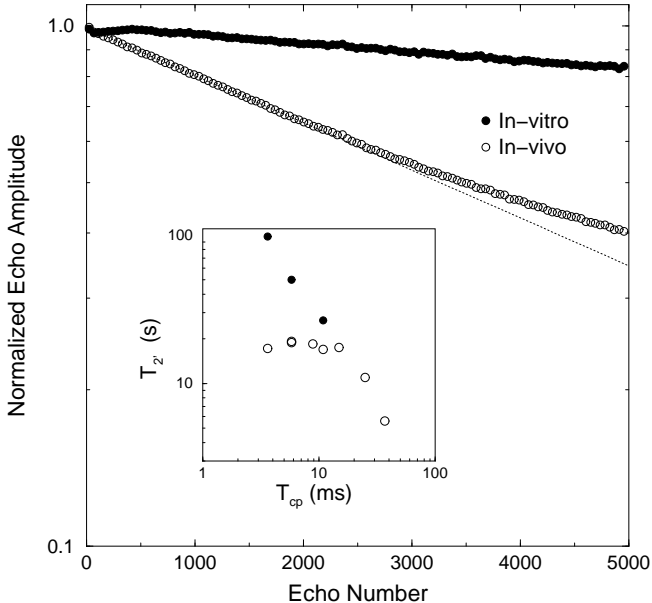}\caption{Fast
repetition CPMG sequence ($T_{\mathrm{CP}}=3.6$~ms) in zero
applied gradient. (For clarity, each symbol represents the average
of 40 consecutive points.) The transverse relaxation rate is much
slower in-vitro (filled
 circles) than in-vivo (open circles). The dotted line indicates the initial, in-vivo,
decay rate; this rate slows in time as the $\mathrm{{O_{2}}}$
concentration decreases. Inset: $T_{2'}$ for various
$T_{\mathrm{CP}}$ in zero applied gradient. At small
$T_{\mathrm{CP}}$, the in-vivo $T_{2'}$ is limited by $T_{1}$
relaxation due to $\mathrm{{O_{2}}}$.}
 \label{fig:pipulse}
\end{figure}

Measurements made in zero applied gradient using a very short
$T_{\mathrm{CP} }=3.6$~ms are shown in Figure~\ref{fig:pipulse}.
For such a fast repetition rate the lock-in amplifier had to be
operated with lower gain which caused a slight decrease in the SNR. The
excellent homogeneity of the $B_{1}$ field allowed for the refocusing of
thousands of echoes with very little loss per RF pulse (see App.~\ref{app1}).
The extremely low power radiated at this frequency allowed these measurements
to be safely performed in-vivo as well.

It is clear from Figure~\ref{fig:pipulse}, that there is marked
difference in decay rate between the in-vitro ($T_{2'}\sim98$~s)
and in-vivo ($T_{2'}\sim 17$~s) results. For $\mathrm{{^{3}He}}$
in the bag, the transverse relaxation at $T_{\mathrm{CP}}=3.6$~ms
would have been limited by the following: diffusion of the gas in
the residual magnet inhomogeneities, $\pi$-pulse losses, or a
combination of both; the longitudinal relaxation, $T_{1}\sim
20$~minutes, was too long to have made a contribution. As shown in
the inset of Figure~\ref{fig:pipulse}, the $T_{2'}$ in the bag
dropped rapidly with increased $T_{\mathrm{CP}}$ indicating that
diffusion was the dominant relaxation mechanism above
$T_{\mathrm{CP}}=3.6$~ms. By contrast, the in-vivo $T_{2'}$
hovered around a much lower value (17~-~19~s for this subject,
16~-~22~s for the other three subjects) before falling off at
$T_{\mathrm{CP}}$ around 15~ms. The $T_{2'}$ values in this
\emph{plateau} region are comparable to the oxygen-limited $T_{1}$
values ($\sim12-20$~s) measured for $\mathrm{{^{3}He}}$ in the
lung by MRI at high field~\cite{deninger1,kauczor}. Complementary
global measurements of $T_{1}$ (decays monitored using small angle
tipping pulses, as described in Ref.~\cite{darrasse})
  and
  $T_{2'}$ (with no applied
gradient) have been performed in-vitro at 2~mT in a different NMR
apparatus. Decay times of several minutes have been  obtained for
small amounts of $\mathrm{{^{3}He}}$ mixed with pure
 $\mathrm{N}_{2}$, while $T_{2'}=T_{1}=$ 13~s were measured for
 $\mathrm{{^{3}He}}$ in air at room pressure and temperature. This strongly
suggests that\ the constant $T_{2'}$ we observed at short
$T_{\mathrm{CP}}$ was indeed due to the nominal $\mathrm{{O_{2}}}$
concentration in the lung. To corroborate this finding, similar
measurements were made where the subject did not inhale air
following a 1-liter bolus of HP gas. In this case, $T_{2'}$ values
greater than 30~s and as high as 36~s were observed (for
$T_{\mathrm{CP}}=5.8$~ms), consistent with a longer $T_{1}$ due to
a lower oxygen concentration.

Another strong signature of $\mathrm{{O_{2}}}$-induced relaxation can be seen
in the main graph of Figure~\ref{fig:pipulse}. The decay rate in-vivo was not
constant, as was seen in-vitro, but instead slowed down over time as the
$\mathrm{{O_{2}}}$ concentration in the lung decreased. This is highlighted by
the dashed line which indicates the initial decay rate of 17~s. A full fit to
the data (not shown) reveals that the echo decay rate, and hence the
$\mathrm{{O_{2}}}$ partial pressure, decreases linearly with time. This has
also been observed by the direct measurement of the $\mathrm{{^{3}He}}$
$T_{1}$ in the lung~\cite{deninger2}.

\subsection{Global Diffusion Measurements}

\begin{figure}[tbh]
\centering
\includegraphics[keepaspectratio,width=3.4in,clip= ]{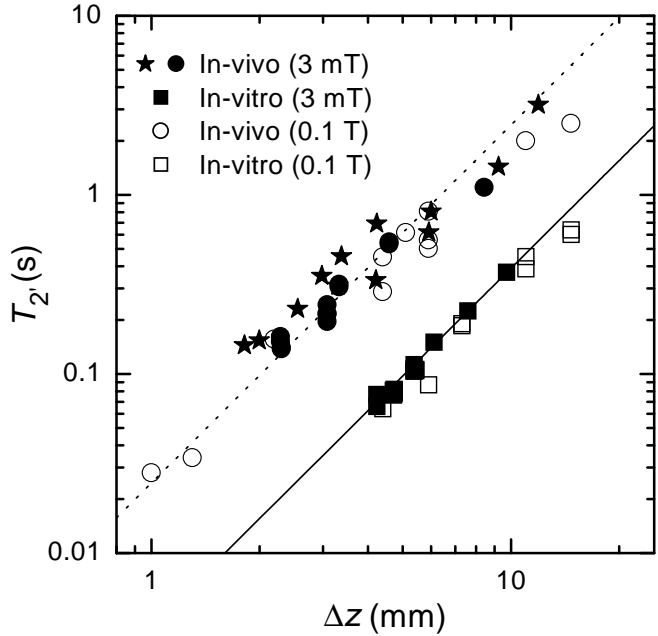}
\caption{Summary of CPMG results plotted as $T_{2'}$ versus an
effective pixel size $\Delta z$ (see text). Data are shown for
in-vivo (circles, stars) and in-vitro (squares) measurements; our
results at 3~mT (closed symbols) are compared to those of
Ref.~\cite{durand} taken at 0.1~T (open symbols). The stars are
the initial 3~mT measurements performed without the Faraday cage.
The solid line is the theoretical response using
$D_{\mathrm{HeN2}}$. The dashed line was generated using
$D_{\mathrm{HeN2}}/6.3$; it acts to guide the eye and to highlight
the restricted diffusion observed in the lung.}
 \label{fig:pixdat}
\end{figure}

A compilation
 of all our diffusion measurements (taken with and
without the Faraday cage, and using two of the subjects only) is
shown in Figure~\ref{fig:pixdat}. These measurements were made
with a minimum $T_{\mathrm{CP}}$ of 8.8~ms. The general behavior
seen here is compared
 with the recently published results of reference~\cite{durand} taken at
0.1~T with a different set of two subjects.
The data is plotted as
$T_{2'}$ versus $\Delta z=2\pi/\left( \gamma
kGT_{\mathrm{CP}}\right)  $, which appears in Eq.~(\ref{eq:tdiff})
as the natural length scale for relaxation due to diffusion.
$\Delta z$ actually differs from the usual pixel size $2\pi/\left(
\gamma G\delta\right)$, which sets the resolution of any MR image
and provides a useful forum to discuss the results, but the two
lengths have similar values in our experiments since
$\delta\approx T_{\mathrm{CP}}$.

It is clear from Figure~\ref{fig:pixdat} that results at 3~mT
(closed symbols) and 0.1~T (open symbols) exhibit the same trend:
for a given value of $\Delta z$, there is a longer $T_{2'}$
in-vivo owing to the restricted diffusion of the
$\mathrm{{^{3}He}}$ in the small airspaces of the lung. This is
fortunate from the point of view of lung imaging, as more time can
be had to obtain the desired image resolution. To extract
diffusion coefficients, noting that
 $1/T_{2}$
  is negligible for
these data, linear fits (through the origin) were made to the data
of Figure~\ref{fig:pixdat} plotted as $\Delta z^{2}$ versus
$T_{2'}$. This procedure gives more weight to measurements with
longer $T_{2'},$ which have a higher SNR, and is thus preferred to
a linear fit of $\Delta z^{-2}$ versus $1/T_{2'}$. For in-vitro
results, we obtained a value of $79.5\pm1~\mathrm{mm^{2}/s}$, in
perfect agreement with the expected free diffusion coefficient
(see App.~\ref{app2}). This is likely to be fortuitous since the
gradient calibration is not accurate to better than 1\% and the
ambient pressure and temperature were not precisely measured
during the experiments. The solid line in Figure~\ref{fig:pixdat}
was drawn using this value to show the theoretical response for
the free diffusion of $\mathrm{{^{3}He}}$ in $\mathrm{{N_{2}}}$.

Using all our in-vivo measurements we obtained a value for the global ADC in
the lung of $14.2\pm0.6~\mathrm{mm^{2}/s}$, whereas results from
reference~\cite{durand} give $22\pm1.5~\mathrm{mm^{2}/s}$ (see Tab.~\ref{tab}%
).\begin{table}[tbh]
\centering%
\begin{tabular}
[c]{c|c|c|}\cline{2-2}\cline{2-3}%
& ADC ($\mathrm{mm^{2}/s}$) & error($\mathrm{mm^{2}/s}$)\\\hline\hline
\multicolumn{1}{|c|}{all 3mT data} & 14.2 & $\pm$0.6\\\hline
\multicolumn{1}{|c|}{0.1T data} & 22 & $\pm$1.5\\\hline\hline
\multicolumn{1}{|c|}{all 3mT data} & 11.9 & $\pm$3.4\\\hline
\multicolumn{1}{|c|}{initial 3mT data} & 11.3 & $\pm$4.2\\\hline
\multicolumn{1}{|c|}{3mT data in cage} & 12.4 & $\pm$2.5\\\hline
\multicolumn{1}{|c|}{0.1T data} & 16 & $\pm$5\\\hline\hline
\end{tabular}
\caption{Table of global ADC values computed from in-vivo CPMG
measurements. Upper two lines: values from global linear fits
($\pm$statistical error) of the data (squares of pixel sizes
versus $T_{2'}$). Lower four lines : values deduced in each
measurement using Eq.~(\ref{eq:tdiff}) (individual ADC values
$\pm$standard deviation).} \label{tab}
\end{table}\ \ \ On average our measurements give a global ADC that is on the
order of$\ D_{\mathrm{HeN2}}/6.3$ (the dashed line in
Figure~\ref{fig:pixdat}). In contrast with in-vitro measurements,
a possible deviation from quadratic dependence and a large scatter
are observed in Figure~\ref{fig:pixdat} for in-vivo measurements.
The small quoted statistical error on the ADC values in the lung
are thus misleading, and it is much more relevant to extract one
ADC value from each measurement and to relate the scatter to
differences between measurements. The corresponding results are
given in the lower part of Table~\ref{tab}. The possible deviation
from quadratic dependence in Figure~\ref{fig:pixdat}
 could not arise from the anisotropy of the lung structure at microscopic length
 scales, as probed using strong bipolar gradients for diffusion
 weighting~\cite{yablonskiy}.
 In this experiment, it would rather indicate the complex
 variation of the gas confinement over the explored length scales
 (2 - 10~mm).  Over the range of diffusion times examined here, the scatter
in the data is likely dominated by differences that
 exist in the physiology of the subjects, and in the degree of lung inflation.
In light of these last two points, it is interesting to note that
 the smallest scatter was obtained for a
  set of measurements that happened to be taken in
a single subject only and where with
 similar breathing maneuvers
were used (data in the Faraday cage, solid circles in
Fig.~\ref{fig:pixdat}).

It is also worthwhile noting  that average ADC values reported
here and in Ref.~\cite{durand} tend to be lower than those of
other healthy subjects that have recently been measured at 1.5~T
using bipolar gradient pulses (17 to $25~\mathrm{mm^{2}/s}$ from
Ref.~\cite{SaamDif} or 16 to $30~\mathrm{mm^{2}/s}$ from
Ref.~\cite{SalernoRad}). Certainly, physiology and imaging
parameters could account for the observed discrepancies, however
one should also consider the difference in probed length scales as
argued in Ref.~\cite{durand}.

\subsection{\label{sec:regional}Regional Diffusion Measurements}

\begin{figure}[tbh]
\centering
\includegraphics[keepaspectratio,width=3.4in,clip= ]{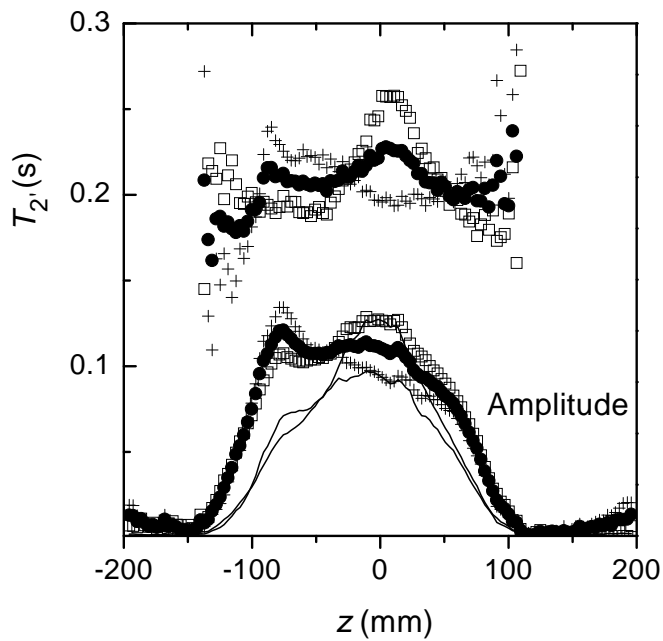}\caption{1D
images of the inital amplitudes (lower traces, arbitrary scale)
and local $T_{2'}$s (upper traces) from in-vivo measurements in
the Faraday cage with opposed applied gradients ($\left|  G\right|
$=660~$\mu$T/m) and otherwise identical conditions
($T_{\mathrm{CP}}$=15.8~ms, $\delta$=13~ms, sampling frequency
10~kHz). Solid lines: uncorrected amplitudes with limited
bandwidth effects. Open squares, crosses: corrected amplitudes and
fitted $T_{2'}$s for the measurements. Filled circles: average
values of the results, which eliminate the main effect of $B_{0}$
inhomogeneities (see text).}
\label{fig:profiles}
\end{figure}

Each recorded echo contains information on the distribution of the polarized
gas along the applied gradient axis (here, the vertical Z axis oriented from
feet to head for a standing subject). FFTs of the spin echoes thus provide
series of 1D images with a nominal resolution given by the pixel size defined
in the previous section. Provided that the SNR is large enough, a pixel by
pixel analysis of the decays can be performed to obtain Z-profiles of the
initial gas distribution and of the local decay time. Examples of results of
such analyses are shown in Figure~\ref{fig:profiles} for two in-vivo
measurements performed with opposite gradient directions. For each measurement, the fitted initial amplitudes (solid lines)
and decay times are obtained from non-linear least squares (NLLS) exponential
fits of the power spectra (i.e. fitting, for each pixel, the squares of the
amplitudes of the echoes by a single exponential decaying to the noise level).
The fitted amplitude profiles are used to compute the actual gas distribution
profiles (symbols) by simply correcting for the acquisition system bandwidth
(in Fig.~\ref{fig:profiles}, the 395 mm field of view (FOV) corresponds to a
frequency range of 10 kHz, while the FWHM bandwidth of the detection coil is
5.6~kHz). The SNR in the central region of the lung with highest signal
intensity is 180 (signal amplitude in Fig.~\ref{fig:profiles} divided by
standard deviation between echoes in similar recordings without polarized
gas). The noise level increases near the edges of the FOV due to the growing
influence of the amplifier noise in the bandwidth-corrected data.

A difference in profiles obtained for opposite directions of the
applied gradient was systematically observed in all experiments,
both in-vivo and in-vitro. It is attributed to the rather strong
magnet inhomogeneities at the position of the measurements, which
result in a non-uniform value of the total gradient. It is much
weaker on the field axis, it scales with $B_{0}$ (as was checked
at 1.5~mT) and it is relatively larger at reduced applied $G$.
Figure~\ref{fig:profiles} clearly shows that an apparent excess
(lack) of signal amplitude is correlated to a longer (shorter)
$T_{2'}$ than the average: both effects result from a locally
decreased (increased) value of the total field gradient. To first
order, the averages of the values obtained from experiments
performed with $\pm G$ (the filled symbols in
Fig.~\ref{fig:profiles}) should thus reproduce the true gas
distribution profile and the true $T_{2'}$ profile which would be
directly measured in a homogeneous $B_{0}$ field.

A clear non-uniform distribution of the average signal is observed
in the high intensity region, however it may result from the
imperfectly compensated effects of the $\pm G$ measurements. The
average $T_{2'}$ profile is quite uniform over most of lung
($\pm5\%$ over the high signal intensity region), and the observed
variations cannot be safely ascribed to actual ADC differences in
spite of the high SNR. Regional dependence of the measured ADC in
the normal human lung has been observed at 1.5~T~\cite{salerno},
with 20\% lower ADC values in the posterior part of the lungs of a
supine lying subject (this is attributed to the known
gravity-induced differences, e.g. of alveolar inflation, in the
normal lung). As is discussed in reference~\cite{durand}, the
length scale involved in such measurements (typically 0.3~mm) is
much smaller than the length scale $\Delta z$ in a CPMG\
measurement, and this may impact on the ADC values.

For all measurements performed in the Faraday cage, the high SNR\
allows to perform a local ADC analysis and to apply the
corrections required to obtain local or global values of the ADC.
However the reduced SNR of all initial measurements (without the
Faraday cage) prevented us from making such corrections, and the
global $T_{2'}$ values in Figure~\ref{fig:pixdat} have been
computed by NLLS exponential fits of the total areas of the power
spectra of the echoes. We have checked on all high SNR data (both
for $+G$ and $-G$) that this procedure actually introduces little
bias ($<$10\%) on the global $T_{2'}$ compared to the true
average, much less in fact that the rather large scatter of the
in-vivo results.

\section{Prospects}

In this paper we have presented NMR measurements of
$\mathrm{{^{3}He}}$ in the human lung made with a simple,
homebuilt system operating at 3~mT. With this device we were able
to demonstrate for the first time some of the proposed benefits of
performing HG imaging at very low field, most notably, human scale
MR with good SNR using a very
 simple scanner design.
 Using a simple wire-wound coil, we were able to generate a
$B_{1}$ field of sufficient homogeneity to observe very long CPMG
decay rates without being obscured by $\pi$-pulse losses or
affected by tissue susceptibility effects. Thanks to the full use
of the available magnetization in a CPMG experiment, this provides
an accurate way to measure the $\mathrm{O}_{2}$ concentration and
its time evolution with small quantities
 of polarized gas. The assessment of the potential interest of
these measurements, and their extension to regional measurements must now be performed.

Good SNR was obtained in the 1D profiles even using modest
quantities of gas and unsophisticated NMR coils. With additional
gradient coils, 2D and 3D imaging is expected to provide results
similar to those already obtained in a commercial 0.1~T
apparatus~\cite{durand}, with the added flexibility allowed by the
ultrafast RF repetition and the very long intrinsic $T_{2}$. This
potential to perform quality lung imaging and accurate ADC mapping
at very low field in a \emph{standing} subject will be fully used
when a dedicated $B_{0}$ field of adequate homogeneity over the
lung volume - a system rather easy to design and build - will be
operational.

\begin{ack}
The authors kindly acknowledge the assistance of V. Senaj for help with
gradient calibrations, and of D. Courtiade, R Labb\'{e} and J.F. Point for
the construction of the Faraday cage. They are grateful for support from the
French Ministry of Research (post doctoral fellowship for
C.B. and grant for project 2000/83) and from the E.C.
FP5  program (PHIL project QLG1-CT-2000-01559).
\end{ack}

\appendix

\section{Field inhomogeneities and tipping angle accuracy}

\label{app1} 
\begin{figure}[tbh]
\centering
\includegraphics[keepaspectratio,width=3.4in,clip= ]{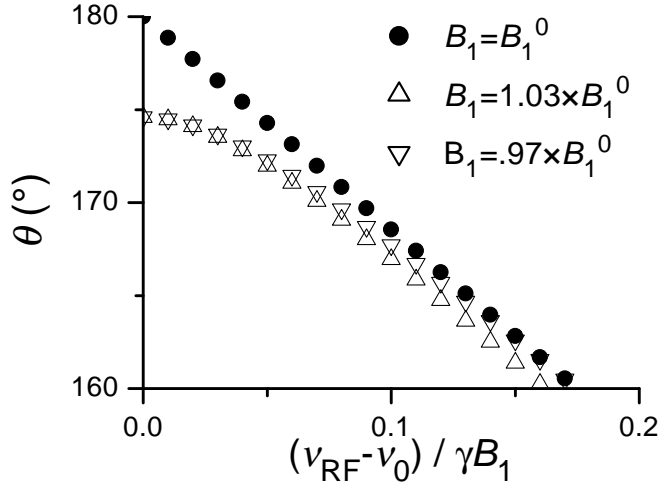}%
\caption{Computed actual tip angle obtained for imperfect 180${{}^{\circ}}$
pulses, due to an off-resonance RF frequency (horizontal axis) and imperfect
RF amplitude (open symbols).}%
\label{fig:app}%
\end{figure}

In the rotating frame synchronous with the near-resonant
component of the RF $\mathbf{B}_{1}$ field, the magnetization tipping at a
given point $\mathbf{r}$ results from the precession during a time
$T_{\mathrm{RF}}$ around the effective field $\mathbf{B}_{\mathrm{eff}%
}=\mathbf{B}_{0}\left(  \nu_{\mathrm{RF}}-\nu_{0}(\mathbf{r})\right)
/\nu_{\mathrm{RF}}+\mathbf{B}_{1}\left(  \mathbf{r}\right)  ,$ where $\nu
_{0}(\mathbf{r})$ is the local Larmor frequency and $\nu_{\mathrm{RF}}$ the
frequency of the $\mathbf{B}_{1}$ field. Noting $\Psi$ the angle of
$\mathbf{B}_{\mathrm{eff}}$ with the Z direction (that of the average
$\mathbf{B}_{0}$ field, see Sec.~\ref{sec:B0B1}), the actual tip angle
$\theta$ at position $\mathbf{r}$ is given by%
\begin{equation}
\cos\theta\left(  \mathbf{r}\right)  =\cos^{2}\Psi+\sin^{2}\Psi\cos\left(
\gamma B_{\mathrm{eff}}T_{\mathrm{RF}}\right)  . \label{eq:theta}%
\end{equation}
180$%
{{}^\circ}%
$ pulses cannot be exactly obtained except at resonance ($\Psi=\pi/2$), and
the largest tip angle which can be obtained is $2\Psi$ for the optimal pulse
duration $T_{\mathrm{RF}}^{\mathrm{opt}}=\pi/\gamma B_{\mathrm{eff}}.$ We now
consider situations where imperfect 180$%
{{}^\circ}%
$ pulses result from small deviations from the resonance conditions and small
spatial variations of $B_{1}.$ Setting the pulse duration to its optimal value
for the average $B_{1}$ field (assuming resonance conditions), the tip angle
computed from \ Eq.~(\ref{eq:theta}) depends on position through the local
values of $B_{0}$ and $B_{1}$ as shown in Figure~\ref{fig:app}.

In our experiment, the RF field is uniform to $\pm$3\%, and $\gamma B_{1}%
=625$~Hz. The $B_{0}$ field is ramped up just before data
acquisition, and the resonance condition at the center of the
apparatus is obtained with a reproducibility and a stability of
$\pm$50~Hz. In the absence of applied gradients, the field
inhomogeneity over a typical lung volume is estimated from FFTs of
spin echoes such as in Figure~\ref{fig:echo} (top panel). The 90
Hz FWHM spread (200 Hz\ spread for 90\% of the spectrum area)
corresponds to $\nu_{\mathrm{RF}}-\nu_{0}(\mathbf{r})=\pm$45~Hz
($\pm $100~Hz). Finally, when switched gradients are used, their
residual value during the RF\ pulses is lower than 0.5\% of the
applied value (at most $\pm$5 kHz over a 395 mm FOV), i.e. $\pm$15
Hz over a lung volume. From the plots in
Figure~\ref{fig:app}, actual angles of 170$%
{{}^\circ} $ (160$ {{}^\circ}$) are expected for detunings of
50~Hz (100~Hz). Since angle errors $\alpha$ are expected to induce
a loss $\alpha^{4}/32$ for each pulse in a CPMG sequence
\cite{cowan}, the corresponding values for this loss are
$2.9\times10^{-5}$ ($4.6\times10^{-4}$). This is consistent with
the negligible contribution from pulse losses inferred in all our
experiments, even when thousands of echoes are recorded as in
Figure~\ref{fig:pipulse}.

\section{Value of $D$ in several gas mixtures}

\label{app2} The free diffusion coefficient $D$ in Eq.~(\ref{eq:tdiff})
depends on the gas temperature $T$, pressure $P$ and composition. For helium,
there is an additional effect of the isotopic nature. Data have been published
for diffusion in helium based on calculations of transport coefficients for
accurate potentials \cite{aziz} and on NMR measurements in $^{3}\mathrm{He}$
\cite{barbe,bock}. Data for binary diffusion of the $^{4}\mathrm{He}%
$-$\mathrm{N}_{2}$ \cite{liner,dunlop79} and $^{4}\mathrm{He}$-$\mathrm{O}%
_{2}$ \cite{dunlop79} systems have been obtained by mass-diffusion
measurements. Using a classical theory for transport in dilute gases
\cite{hirsch}, we provide numerical formulas to compute $D$ for a
$^{3}\mathrm{He}$ gas with partial pressure $P_{3}$ in a gas containing other
components ($^{4}\mathrm{He}$, $\mathrm{N}_{2}$, $\mathrm{O}_{2}$) with
partial pressure $P_{4}$, $P_{\mathrm{N2}}$ and $P_{\mathrm{O2}}$, in the
simple case of uniform temperature and partial pressures.

We first evaluate the self- and binary-diffusion coefficients of
$^{3}\mathrm{He}$, $D_{3}$, $D_{34}$, $D_{3\mathrm{N2}}$ and $D_{3\mathrm{O2}%
}$. With notations of Reference~\cite{hirsch},
\begin{equation}
D_{3}=\frac{3}{16}\frac{\sqrt{2\pi\left(  k_{B}T\right)  ^{3}/m_{3}}}{P_{3}%
\pi\sigma^{2}\Omega^{\left(  1,1\right)  \ast}}\label{eq:D3}%
\end{equation}
where $k_{B}$ is the Boltzmann constant, $m_{3}$ the
$^{3}\mathrm{He}$ atomic mass, $\pi\sigma^{2}$ the collision cross
section in a rigid-sphere model and $\Omega^{\left(  1,1\right)
\ast}$ a $T$-dependent factor depending on the actual interatomic
potential. The pressure-independent reduced diffusion coefficient
$\mathcal{D}_{3}=P_{3}D_{3}$ has an apparent $T^{3/2}$ temperature
dependence (actually assumed in Ref.~\cite{bock}), but this is not
exact for a real gas. For helium
($^{3}\mathrm{He,}^{4}\mathrm{He}$ and isotope mixtures), the
computed values in the 200-600~K range \cite{aziz} can be
accurately reproduced (within 0.2\%) by a power law
$\mathcal{D}_{i}\left(  T\right) \propto T^{1.71}.$ For a binary
mixture in which $^{3}\mathrm{He}$ (with negligible partial
pressure) diffuses in another component $i,$ $D_{3i}$ is given by
Eq.~(\ref{eq:D3}) in which the $^{3}\mathrm{He}$ mass is replaced
by the reduced mass $\mu_{3i}=m_{3}m_{i}/\left( m_{3}+m_{i}\right)
,$ $P_{3}$ by $P_{i}$ (the foreign gas pressure) and
$\Omega^{\left(  1,1\right)  \ast}$ depends on the involved
interaction potential. The reduced diffusion coefficient
$\mathcal{D}_{3i}=P_{i}D_{3i}$ again only depends on temperature.
For $\mathrm{N}_{2}$, the values measured in the 300-800~K range
\cite{liner} can be nicely fit by
$\mathcal{D}_{4\mathrm{N2}}\left(  T\right)  \propto T^{1.65}.$
For both $\mathrm{N}_{2}$ and $\mathrm{O}_{2},$ the ratios of the
measurements of reference~\cite{dunlop79} provide the same
exponent 1.68 but this determination is less accurate due to the
small temperature interval (300 and 323~K). We thus choose to use
the same exponent 1.65 for $\mathrm{N}_{2}$ and $\mathrm{O}_{2}.$\
All these experimental data have been obtained for
$^{4}\mathrm{He}$, and the faster diffusion of $^{3}\mathrm{He}$
can be evaluated using the ratio of the relevant reduced masses.
This procedure provides the correct
  values within 0.1\% compared to the diffusion of
helium~\cite{aziz} or helium-hydrogen~\cite{dunlop95} isotope mixtures. This
results from the negligible effect of quantum statistics on the binary
collisions at room temperature, and is even more accurate for collisions with
a heavier molecule. These isotope effects and the numerical formulas for
$\mathcal{D}_{3}(T)$ and $\mathcal{D}_{3i}(T)$ are collected in
Table~\ref{tab2}. \begin{table}[tbh]
\centering%
\begin{tabular}
[c]{|c|}\hline
$D_{3}/D_{4}=\sqrt{4/3}$\\\hline
$D_{34}/D_{4}=\sqrt{7/6}$\\\hline
$D_{3N2}/D_{4N2}=\sqrt{31/24}$\\\hline
$D_{3O2}/D_{4O2}=\sqrt{35/27}$\\\hline
$\mathcal{D}_{3}=P_{3}D_{3}=1.997\times\left(  T/300\right)  ^{1.71}$\\\hline
$\mathcal{D}_{34}=P_{4}D_{34}=1.868\times\left(  T/300\right)  ^{1.71}$\\\hline
$\mathcal{D}_{3N2}=P_{N2}D_{3N2}=0.811\times\left(  T/300\right)  ^{1.65}$\\\hline
$\mathcal{D}_{3O2}=P_{O2}D_{3O2}=0.857\times\left(  T/300\right)  ^{1.65}$\\\hline
\end{tabular}
\caption{First 4 lines: isotope effect on the self- or binary-diffusion
coefficients of helium at given temperature and pressure. Last 4 lines:
numerical expressions to compute the reduced diffusion coefficients
$\mathcal{D}(T),$ in units of $\mathrm{atm}\times\mathrm{cm}^{2}\mathrm{/s}$
(1 atm = $1.013\times10^{5}$ Pa), with the temperature $T$ in Kelvin.}%
\label{tab2}%
\end{table}

The free diffusion coefficient $D$ in a gas mixture is finally given from
these reduced diffusion coefficients by :%
\begin{equation}
\frac{1}{D}=\frac{P_{3}}{\mathcal{D}_{3}(T)}+\frac{P_{4}}{\mathcal{D}_{34}%
(T)}+\frac{P_{\mathrm{N2}}}{\mathcal{D}_{3\mathrm{N2}}(T)}+\frac
{P_{\mathrm{O2}}}{\mathcal{D}_{3\mathrm{O2}}(T)}.\label{eq:Dtot}%
\end{equation}

Eq.~(\ref{eq:Dtot}) can be applied for instance to the gas mixtures used for
the in-vitro measurements of Figure~\ref{fig:pixdat}, for which typically 30
cm$^{3}$ of $^{3}\mathrm{He}$ were mixed to 1 liter of $\mathrm{N}_{2}$: for 1
atm at 20$%
{{}^\circ}%
\mathrm{C,}$ one obtains $D=79.5$~mm$^{2}$/s, slightly larger than
$D_{3\mathrm{N2}}=78.1$~mm$^{2}$/s. Eq.~(\ref{eq:Dtot}) can also be applied to
evaluate the \emph{free} diffusion coefficient for typical in-vivo
experiments, in which the inhaled dose (30 cm$^{3}$ of $^{3}\mathrm{He}$ mixed
with 0.5 liter of $\mathrm{N}_{2}$) mixes with the subject's lung contents
($\sim$5 liters of $\mathrm{O}_{2}$-depleted air, since the typical partial
pressure of $\mathrm{O}_{2}$ is 0.13 atm in-vivo instead of 0.21 atm in room
air). For 1 atm of this mixture at 37$%
{{}^\circ}%
\mathrm{C,}$ one obtains $D=86.6$ mm$^{2}$/s. This higher value is mostly due
to the higher temperature, and to a lesser extent to the faster diffusion of
helium in $\mathrm{O}_{2}.$

\end{document}